\documentclass[prb,preprint]{revtex4-1} 

\usepackage{amsmath}
\usepackage{amsfonts} 
\usepackage{graphicx} 
\usepackage{multirow} 

\usepackage{tabu}
\graphicspath{ {images/} }

\begin{document}
\title{A purely geometrical  method of determining the location of a smartphone accelerometer}

\author{Christopher Isaac Larnder}\email{chrisisaac.larnder@johnabbott.qc.ca}
\affiliation{Department of Physics, John Abbott College, St-Anne-de-Bellevue QC, Canada H9X 3X8}

\begin{abstract}
In a paper ( posthumously ) co-authored by Isaac Newton himself\cite{Newton}, 
the primacy of geometric notions in pedagogical expositions of centripetal acceleration has been clearly asserted. 
In the present paper we demonstrate how this pedagogical prerogative can 
inform the design of
an experiment involving an accelerometer-equipped smartphone rotating uniformly in a horizontal plane. Specifically, the location of the sensor itself within the body of the smartphone will be determined using a technique that is purely geometrical in nature, relying on nothing more than the notion that centripetal accelerations are centrally-pointing.
The complete absence of algebraic manipulations obliges students to focus exclusively on the development of their geometrical reasoning abilities.
In particular, it provides a healthy challenge for
those algebraically-accomplished students for whom equations, calculations and data tables represent a means of avoiding a direct confrontation with the imposing spectre of material that is otherwise purely conceptual in nature.
\end{abstract}

\maketitle

\section{Introduction}

A recent paper\cite{LarnderLarade} determined accelerometer positions across a wide range of host devices including not only smartphones but also tablets and dedicated accelerometry devices. The authors applied a simple but rigorous mathematical technique relying on the vector form of the equation for the centripetal acceleration $\vec{a}$, viz.

\begin{equation}
\label{eqVector}
\vec{a} = -\omega^2 \vec{R}, 
\end{equation}
in which $\vec{R}$ is the position vector of the sensor with respect to a coordinate system whose origin coincides with the pivot point.  Its determination relies on the values of $\vec{a}$ and of the angular velocity $\omega$. Earlier attempts
\cite{monteiro1}\cite{brasileiro}\cite{monteiro2} \cite{vogt}\cite{hochberg}\cite{vieyra} \cite{mau}, reviewed in the same paper, are more geometrical in nature, but still rely on the numerical value of $\omega$ and a computation based on the scalar form of the same equation, viz.

\begin{equation}
\label{eqScalar}
 a= \omega^2 R.
\end{equation}
Rewriting Eq. \ref{eqVector} as

\begin{equation}
\label{eqBoth}
a\hat{a}= -\omega^2 R\hat{R}
\end{equation}
demonstrates its separability into two equations, one involving only magnitudes (Eq. \ref{eqScalar}) and
 the other involving purely geometrical quantities expressed as unit vectors, viz.

\begin{equation}
\label{eqDirections}
\hat{a}= -\hat{R}
\end{equation}

Since the position vector $\vec{R}$ points from the pivot point to the sensor, the acceleration vector points from the sensor back towards the pivot point: It is a centrally-pointing vector. Whereas previous work has relied on Eq. \ref{eqVector} or Eq. \ref{eqScalar}, we rely only on the purely geometric relationship expressed in Eq. \ref{eqDirections}.

For most students this introductory discussion should be avoided altogether, as it pertains mostly to notation. Their attention should be exclusively occupied with the centrally-pointing property of the acceleration vector and its implications in the context of the experiment. 

\section{Methods and Results}

\begin{figure} [h]
\includegraphics[width=14 cm]{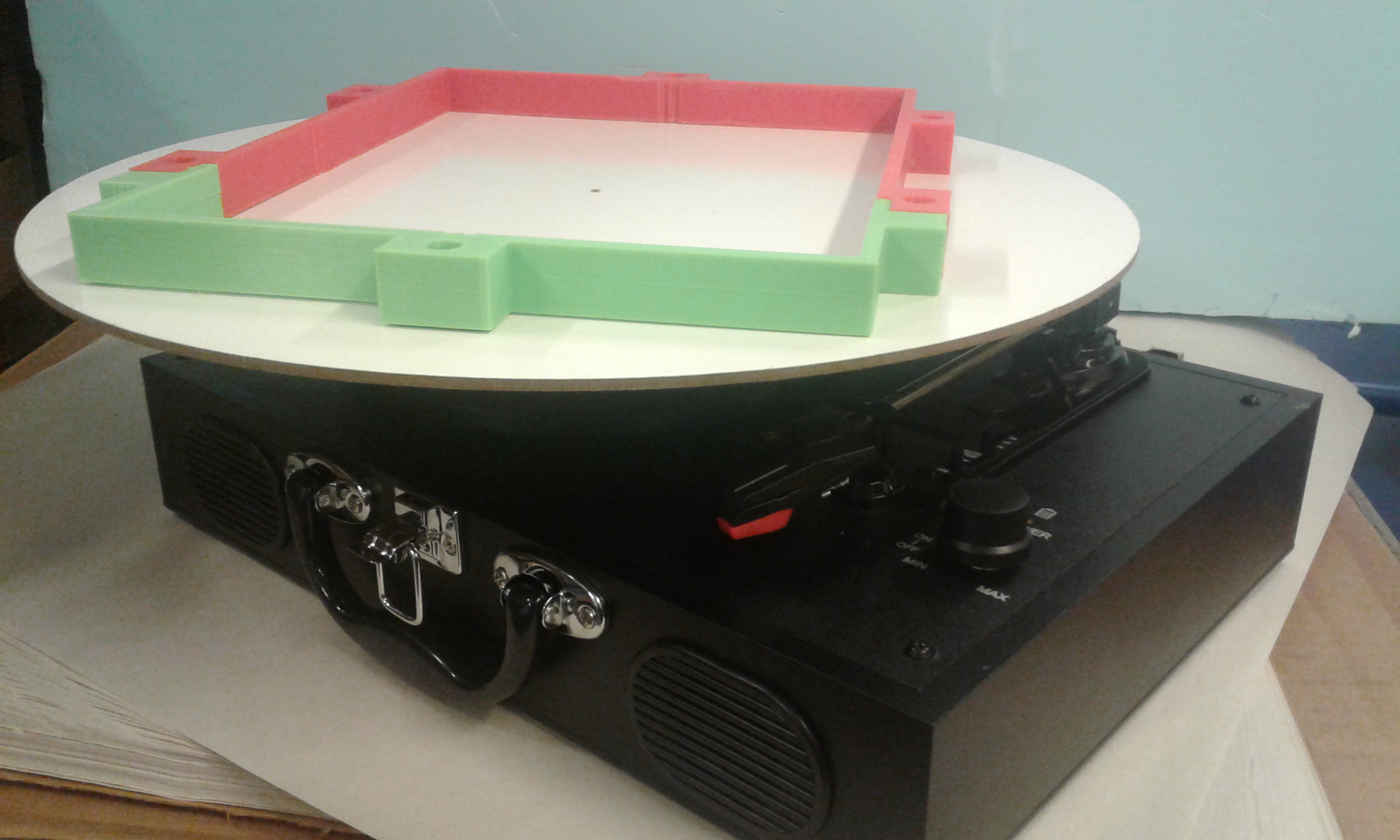}
\centering
\caption{The ``SpinFrame'' apparatus: a 3D-printed frame admitting an 8-1/2-by-11 paper is connected to a standard turntable via a 3D-printed fitting ( the fitting is not visible here). }
\label{fig:apparatus}
\end{figure}

The apparatus, a photo of which is presented in Fig. \ref{fig:apparatus}, consists of a 3D-printed rectangular frame mounted on a flat disk, which in turn is mounted onto a turntable\cite{turntable} using an additional 3D-printed fitting
\cite{3D}
. The dimensions of the frame are such that a standard 8.5 x 11 - inch piece of paper fits snugly within it.
Students are expected to arrive in class with an accelerometer app already downloaded onto their smartphones\cite{TheApp}. They initiate a recording with the app, place it in one of the four corners, and set the frame spinning at a constant angular velocity\cite{rate}. The data is transferred to a PC and average values for the acceleration components are obtained.

\begin{figure} [h]
\includegraphics[width=14 cm]{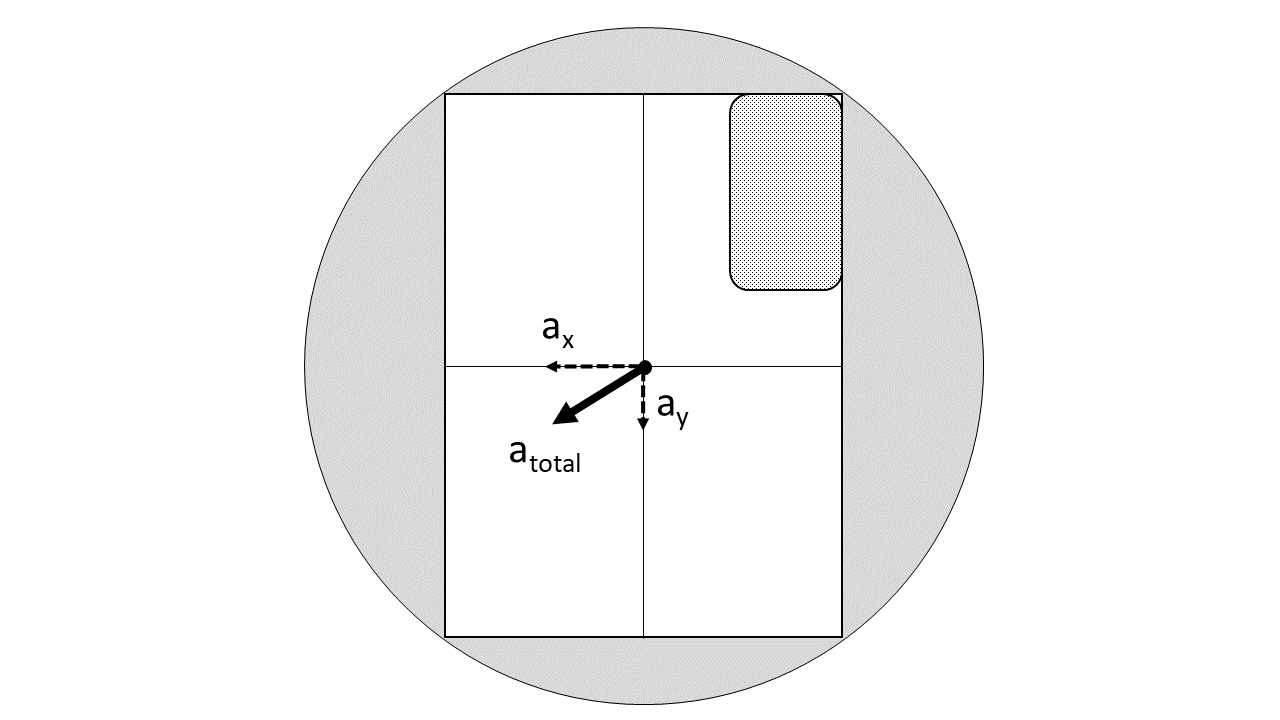}
\centering
\caption{Reconstruction of the acceleration vector from its components for a smartphone placed in the first-quadrant corner.}
\label{fig:Slide1}
\end{figure}

\begin{figure} [h]
\includegraphics[width=14 cm]{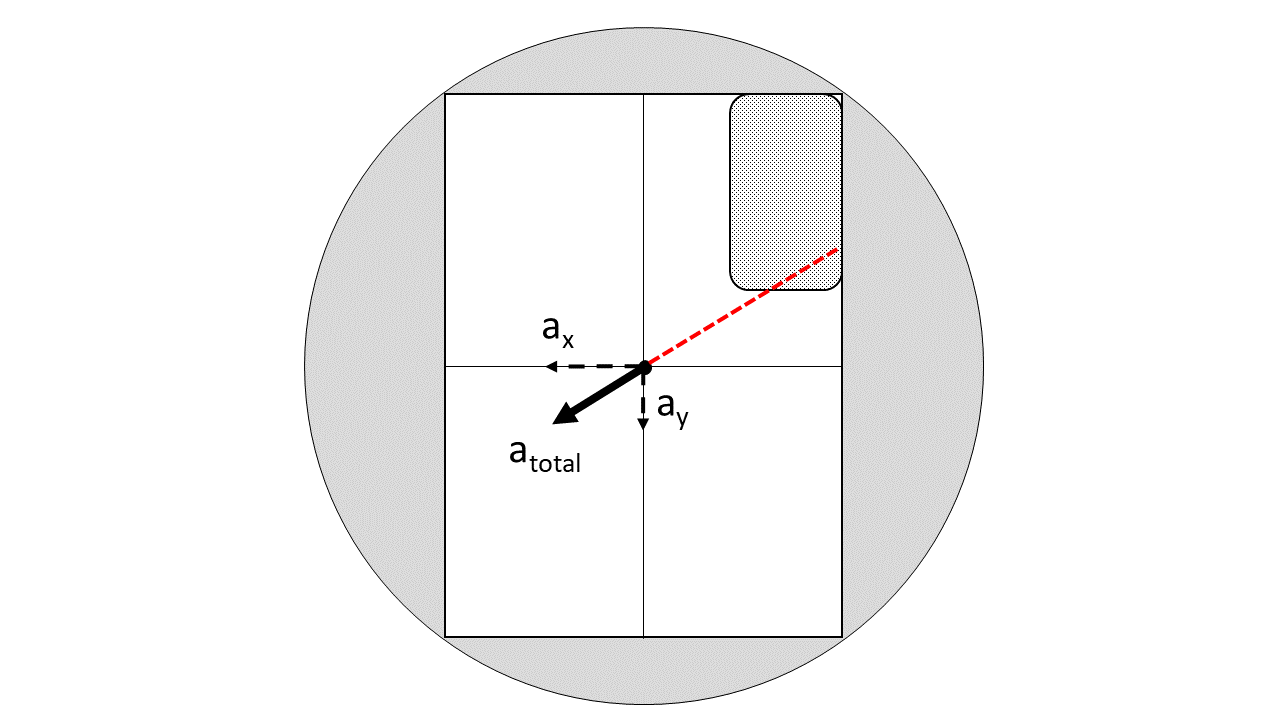}
\centering
\caption{Tracing of the radial line associated with the current acceleration vector.}
\label{fig:Slide2}
\end{figure}

The two acceleration components are drawn on the frame paper, from the center outwards, and used to form a vector indicating the direction of acceleration, as depicted in Fig. \ref{fig:Slide1}. For a turntable at 78 rpm, a scale factor of 
1.0 m/s\textsuperscript{2} = 1 cm will ensure all drawn vectors fit on the sheet of paper.
This is a good point to remind students that the acceleration vector, when drawn from the true (but as-yet unknown) sensor position outwards, must i) point towards the center and ii) have the same angle as the vector they have just drawn. With these points in mind\cite{PreLab}, have them try to identify positions at which the sensor could have been located.
These constitute candidates for the true position of the accelerometer sensor within the body of the phone. The astute student will realize that there are multiple such points, that the collection of all such points forms a line, and that an extension of this line will intersect the origin: it is a 
radial line and is the unique radial line associated with the given acceleration vector.
It follows immediately that this radial line can be established in a rigorous manner by simply extending the acceleration vector backward through the area in which the phone had been placed, as illustrated in Fig. \ref{fig:Slide2}. 

\begin{figure} [h]
\includegraphics[width=14 cm]{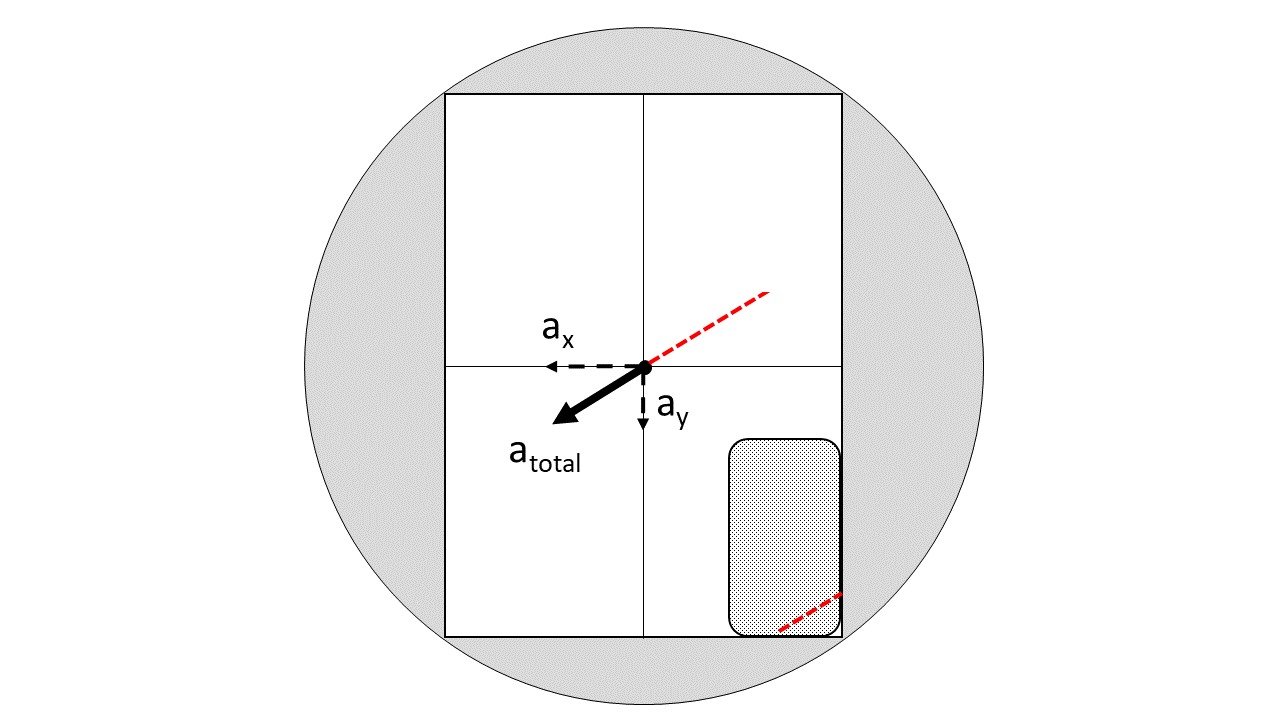}
\centering
\caption{The paper outline, with the radial line traced over it, is moved from quadrant 1 to quadrant 4.}
\label{fig:Slide3}
\end{figure}

\begin{figure} [h]
\includegraphics[width=14 cm]{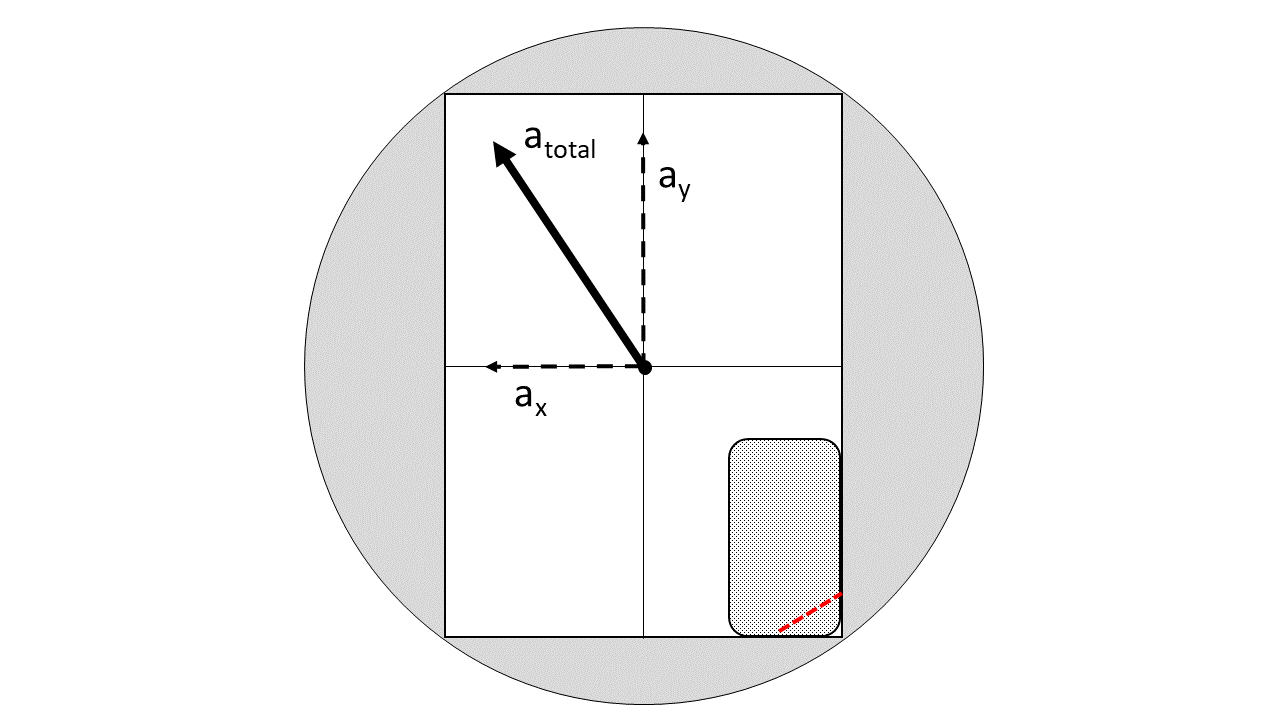}
\centering
\caption{Reconstruction of the acceleration vector for quadrant 4.}
\label{fig:Slide4}
\end{figure}

\begin{figure} [h]
\includegraphics[width=14 cm]{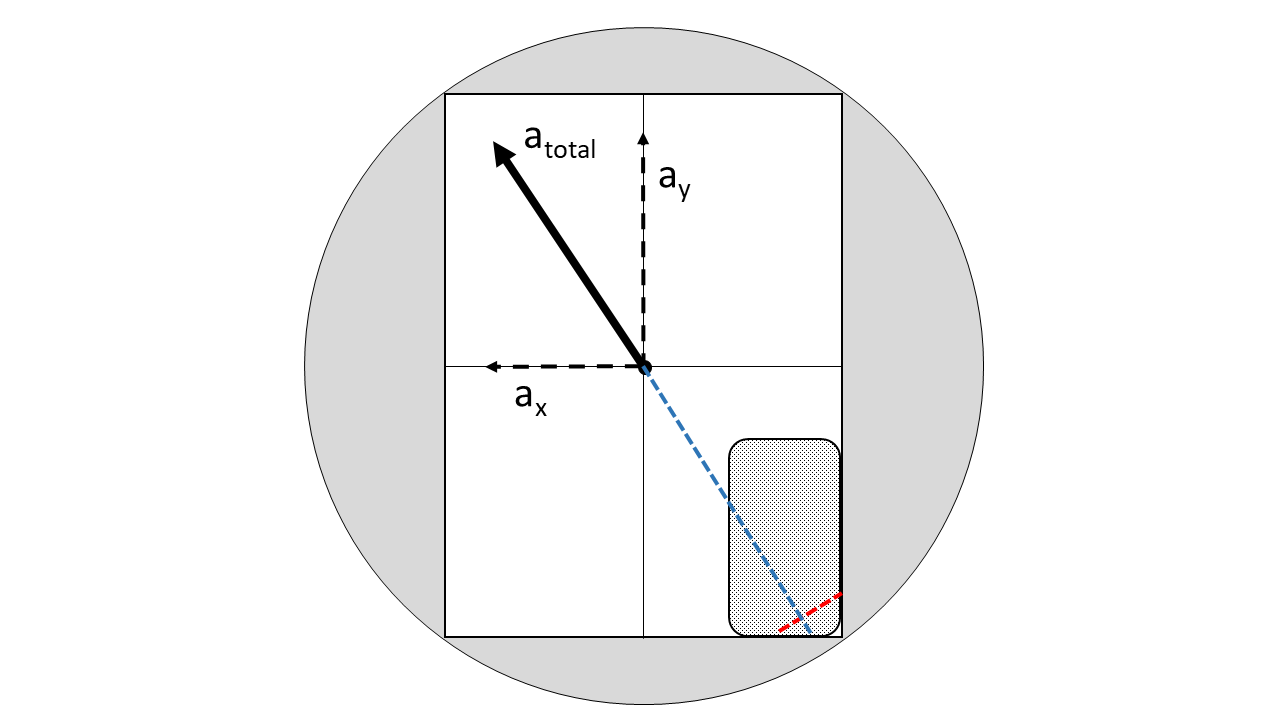}
\centering
\caption{Tracing of the radial line for quadrant 4. The superposition of the two radial lines provides an initial estimate for the true sensor position.}
\label{fig:Slide5}
\end{figure}

\begin{figure} [h]
\includegraphics[width=14 cm]{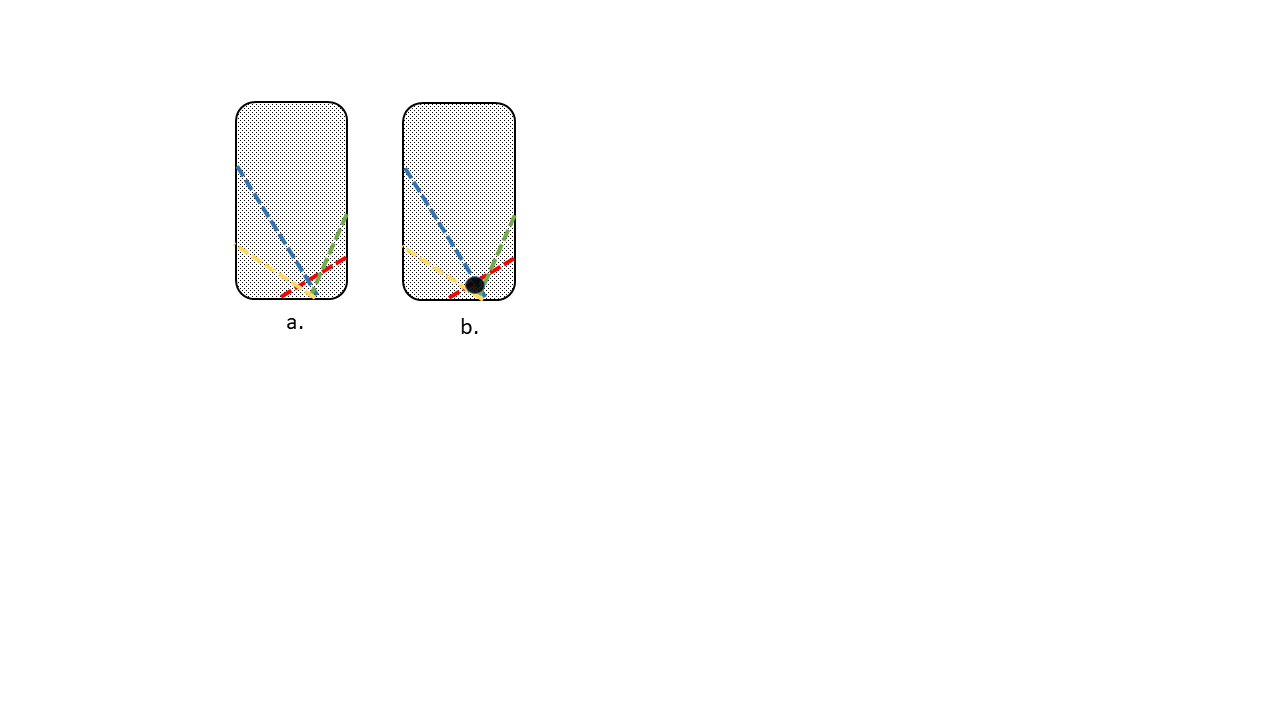}
\centering
\caption{a: Result of accumulating radial lines from each of the four quadrants. b: Best-estimate position of the sensor.}
\label{fig:Slide6}
\end{figure}

The smartphone is then moved consecutively to each of the other quadrants and the experiment is repeated.
The students, having beforehand traced and cut out a rectangular outline of their phone, place this outline at each of the corners and trace each radial line over it. Figs.\ref{fig:Slide3}, \ref{fig:Slide4} and \ref{fig:Slide5} illustrate one step in this cumulative procedure. The accumulation of 4 radial lines, as in Fig. \ref{fig:Slide6}, provides a means of visually estimating both a position and an uncertainty in the position value.

\begin{figure} [h]
\includegraphics[width=14 cm]{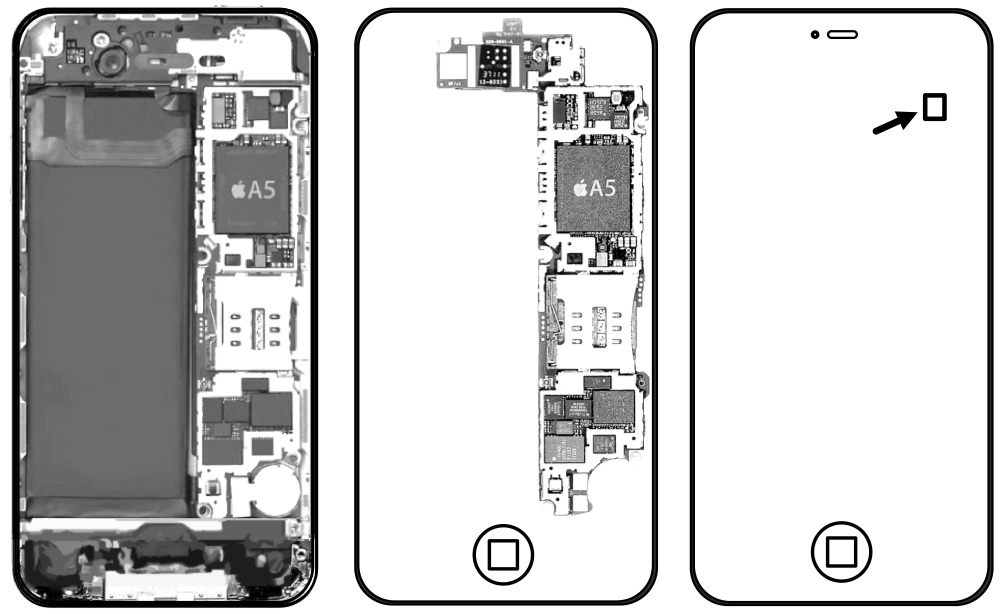}
\centering
\caption{Example of a circuit diagram superimposed on a device outline, used for obtaining a direct estimate of accelerometer position. The host device in this case was an iPhone 4S.}
\label{fig:circuit}
\end{figure}

The only way of directly verifying the accuracy of the result would be to invite students to dismantle their smartphones and locate the sensor on the circuit board. Given the general reluctance to take the spirit of scientific inquiry to such a level, we suggest a lower-risk alternative that involves browsing for the smartphone model on "teardown" sites\cite{teardown} that document the dismantling process and identify the components. Fig. \ref{fig:circuit} depicts a circuit board obtained from such a site, scaled to fit appropriately into the smartphone outline.

\section{Conclusions}
The procedure, as advertised, relies entirely on geometrical reasoning. We do not rely on the value of $\omega$ nor on the ratio of the acceleration components. There is no calculation of angles using the inverse tangent function. Even the scale ratio, involving a 1:1 relationship, is chosen so as to eliminate the need for calculations. Thus freed of a category of activity that often occupies a considerable portion of the time spent in lab, 
teachers can afford to structure the
lab activity in such a way as to lead students towards discovering some of the procedural steps themselves\cite{PreLab}.

More importantly, the absence of numeric activities means that
students are left with nothing to contemplate but the radially-inward lines that they themselves have constructed from the acceleration components of their own phone. They have nothing to gain from a consideration of the algebraic relationship of (scalar) Eq. \ref{eqScalar} and everything to gain from the geometric relationship of ( unit-vector) Eq. \ref{eqDirections}. We believe such an approach produces favorable conditions for the emergence of the geometric insight that is so critical to an appreciation of centripetal acceleration.

\section{Acknowledgements}

The author thanks colleagues Brian Larade, Etienne Portelance, Margaret Livingstone, Bruce Tracy  and Hubert Camirand for helpful suggestions and in-class evaluations of this lab activity; and the Engineering Technologies department at John Abbott College for extensive 3D-print technical support. This work was funded in part by Qu\'{e}bec's \textit{Minist\`{e}re de l'\'{E}ducation et de l'Enseignement sup\'{e}rieure}, through contributions from the Canada-Qu\'{e}bec Agreement on Minority-Language Education and Second-Language Instruction.

\begin{thebibliography}{5}
\bibitem{3D}
The apparatus has been baptized as the ``SpinFrame''. The 3D files and assembly instructions have been released under a
CreativeCommons (CC-BY-NC) license and are available
from the author upon request.
\bibitem{rate}
We found an angular rate of 78 rpm to provide data with a good signal-to-noise ratio.
\bibitem{TheApp}
The accelerometer app that we used is called "Physics Toolbox Accelerometer".
\bibitem{PreLab}
We have designed a set of pre-lab exercises which lead students towards this type of reasoning without, in our opinion, being overly prescriptive. They have been released under a CreativeCommons (CC) license and are available from the author upon request.
\end{thebibliography}
\end{document}